\documentclass[11pt,a4paper]{article}
\usepackage[hyperref]{emnlp2020}
\usepackage{times}
\usepackage{latexsym}

\usepackage[flushleft]{threeparttable}
\usepackage[linesnumbered,ruled,vlined]{algorithm2e}
\usepackage{algpseudocode}
\usepackage{booktabs}
\usepackage{subcaption}
\usepackage[nointegrals]{wasysym}
\usepackage{tikz}
\usepackage{tabu}
\usepackage{tabularx}
\usepackage{float}
\usepackage{listings}
\usepackage{color}
\usepackage{multirow}
\usepackage{framed}
\usepackage{pifont}
\usepackage{dsfont}
\usepackage{xcolor}
\usepackage{amsmath,amsthm}
\usepackage{amssymb}
\usepackage{graphicx}
\usepackage{textcomp}
\usepackage{tcolorbox}
\usepackage{soul}
\usepackage{makecell}
\usepackage{xspace}

\definecolor{DarkRed}{RGB}{130,25,0}

\newcommand{\crowdaq}{\textsc{Crowdaq}\xspace}
\newcommand{\ignore}[1]{}
\definecolor{mypurple}{RGB}{121,55,196}
\definecolor{myblue}{RGB}{60,177,245}
\definecolor{myorange}{RGB}{243,206,84}
\definecolor{mygreen}{RGB}{41,222,35}

\newcounter{exctr}

\newcounter{eventCtr}

\newcounter{timexCtr}

\usepackage{listings}
\definecolor{dkgreen}{rgb}{0,0.6,0}
\definecolor{gray}{rgb}{0.5,0.5,0.5}
\definecolor{mauve}{rgb}{0.58,0,0.82}

\usepackage{paralist}
\newcommand{\qn}[1]{{\color{black} {#1}}}

\usepackage{microtype}

\aclfinalcopy 

\title{Easy, Reproducible and Quality-Controlled Data Collection \\with \crowdaq{}}

\author{\makecell{Qiang Ning$^\spadesuit$ ~ Hao Wu$^{\clubsuit}$ ~ Pradeep Dasigi$^{\spadesuit}$ ~ Dheeru Dua$^{\diamondsuit}$~Matt Gardner$^{\spadesuit}$\\
Robert L. Logan IV$^{\diamondsuit}$ ~ Ana Marasovi\'{c}$^{\spadesuit}$ ~ Zhen Nie$^{\clubsuit}$}\\
  $^{\spadesuit}$Allen Institute for AI \quad
  $^{\diamondsuit}$University of California, Irvine \quad
  $^{\clubsuit}$Hooray Data Co., Ltd\\
  {\tt \{qiangn,pradeepd,mattg,anam\}@allenai.org}\\
  {\tt \{haowu,zhennie\}@hooray.ai}\\
  {\tt \{ddua,rlogan\}@uci.edu}
}

\begin{document}
\maketitle
\begin{abstract}
High-quality and large-scale data are key to success for AI systems. However, large-scale data annotation efforts are often confronted with a set of common challenges: (1) designing a user-friendly annotation interface; (2) training enough annotators efficiently; and (3) reproducibility. To address these problems, we introduce \crowdaq{},\footnote{\underline{Crowd}sourcing with \underline{A}utomated \underline{Q}ualifcation; \url{https://www.crowdaq.com/}} an open-source platform that standardizes the data collection pipeline with customizable user-interface components, automated annotator qualification, and saved pipelines in a re-usable format. We show that \crowdaq{} simplifies data annotation significantly on a diverse set of data collection use cases and we hope it will be a convenient tool for the community.

\end{abstract}

\section{Introduction}
Data is the foundation of training and evaluating AI systems.
Efficient data collection is thus important for advancing research and building time-sensitive applications.\footnote{This holds not only for collecting static data annotations, but also for collecting human judgments of system outputs.}
Data collection projects typically require many annotators working independently to achieve sufficient scale, either in dataset size or collection time.
To work with multiple annotators, data requesters (i.e., AI researchers and engineers) usually need to design a user-friendly annotation interface and a quality control mechanism. 
However, this involves a lot of overhead: we often spend most of the time resolving frontend bugs and manually checking or communicating with individual annotators to filter out those who are unqualified, instead of focusing on core research questions.

Another issue that has recently gained more attention is reproducibility. \citet{DGCSS19} and \citet{pineau_reproducibility} provide suggestions for {\em system} reproducibility, and \citet{bender-friedman-2018-data} and \citet{Gebru2018DatasheetsFD} propose ``data statements'' and ``datasheets for datasets'' for {\em data collection} reproducibility.
However, due to irreproducible human interventions in training and selecting annotators and the potential difficulty in replicating the annotation interfaces, it is 
often difficult to reuse or extend an existing data collection project.


We introduce \crowdaq{}, an open-source data annotation platform for NLP
research designed to minimize overhead and improve reproducibility. It has the following contributions.
First, \crowdaq{} standardizes the design of data collection pipelines, and separates that from software implementation.
This standardization allows requesters to design data collection pipelines {\em declaratively} without being worried about many engineering details, which is key to solving the aforementioned problems (Sec.~\ref{sec:pipeline}).

Second, \crowdaq{} automates qualification control via multiple-choice exams. We also provide detailed reports on these exams so that requesters know how well annotators are doing and can adjust bad \qn{exam} questions if needed (Sec.~\ref{sec:pipeline}).

Third, \crowdaq{} carefully defines a suite of pre-built UI components that one can use to compose complex annotation user-interfaces (UIs) for a wide variety of NLP tasks without expertise in HTML/CSS/JavaScript (Sec. \ref{sec:interface}). For non-experts on frontend design, \crowdaq{} can greatly improve efficiency in developing these projects.

Fourth, a dataset collected via \crowdaq{} \qn{can} be more easily reproduced or extended by {\em future \qn{data} requesters},
because they can simply copy the pipeline and pay for additional annotations, or treat existing pipeline as a starting point for new projects.

In addition, \crowdaq{} has also integrated many useful features: requesters can conveniently monitor the progress of annotation jobs, whether they are paying annotators fairly, and the agreement level of different annotators on \crowdaq{}.
Finally, Sec.~\ref{sec:usage} shows how to use \crowdaq{} and Amazon Mechanical Turk (MTurk)\footnote{\url{https://www.mturk.com/}} to collect data for an example project.
More use cases can be found in our documentation.

\section{Standardized Data Collection Pipeline}
\label{sec:pipeline}
A data collection project with multiple annotators generally includes some or all of the following:
(1) Task definition, which describes what should be annotated.
(2) Examples, which enhances annotators' understanding of the task.
(3) Qualification, which tests annotators' understanding of the task and only those qualified can continue; this step is very important for reducing unqualified 
annotators.
(4) Main annotation process, where qualified annotators work on the task.
\crowdaq provides easy-to-use functionality for each of these components of the data collection pipeline, which we expand next.

\paragraph{\textsc{Instruction}} A Markdown document that defines a task and instructs annotators how to complete the task.
It supports various formatting options, including images and videos. 

\paragraph{\textsc{Tutorial}} Additional training material provided in the form of multiple-choice questions with provided answers that workers can use to gauge their understanding of the \textsc{Instruction}. 
\crowdaq{} received many messages from real annotators saying that \textsc{Tutorials} are quite helpful for learning tasks.

\paragraph{\textsc{Exam}} A collection of multiple-choice questions similar to \textsc{Tutorial}, but for which answers are not provided to participants.  \textsc{Exam} is used to test whether an annotator understands the instructions sufficiently to provide useful annotations.
Participants will only have a finite number of opportunities specified by the requesters to work on an \textsc{Exam}, and each time they will see a random subset of all the exam questions. 
After finishing an \textsc{Exam}, participants are informed of how many mistakes they have made and whether they have passed, but they do not receive feedback on individual questions. Therefore, data requesters should try to design better \textsc{Instructions} and \textsc{Tutorials} instead of using \textsc{Exam} to teach annotators.

We restrict \textsc{Tutorials} and \textsc{Exams} to always be in a multiple-choice format, irrespective of the original task format, because it is natural for humans to learn and to be tested in a discriminative setting.\footnote{E.g., we can always test one's understanding of a concept by multiple-choice questions like {\em Do you think something is correct?} or {\em Choose the correct option(s) from below.}}
An important benefit of using multiple-choice questions is that their evaluation can be automated easily, minimizing the effort a requester spends on manual inspections. Another convenient feature of \crowdaq{} is that it displays useful statistics to requesters, such as the distribution of scores in each exam and which questions annotators often make mistakes on, which can highlight areas of improvement in the \textsc{Instruction} and \textsc{Tutorial}. Below is the JSON syntax to specify \textsc{Tutorials/Exams} (see Fig.~\ref{fig:tutorial} and Fig.~\ref{fig:exam} in the appendix). 
\begin{lstlisting}
"question_set": [
    {
        "type": "multiple-choice",
        "question_id": ...,
        "context": [{
            "type": "text",
            "text": "As of Tuesday, 144 of the state's then-294 deaths involved nursing homes or longterm care facilities."
        }],
        "question": {
            "question_text": "In \"294 deaths\", what should you label as the quantity?",
            "options": {"A": "294", "B": "294 deaths"}
        },
        "answer": "A",
        "explanation": {
            "A": "Correct",
            "B": "In our definition, the quantity should be \"294\"."
        }
    },
    ...
]
\end{lstlisting}

\paragraph{\textsc{Task}} 
For example, if we are doing sentence-level sentiment analysis, then a \textsc{Task} is to display a specific sentence and require the annotator to provide a label for its sentiment.
A collection of \textsc{Tasks} are bundled into a \textsc{Task Set} that we can launch as a group.
Unlike \textsc{Tutorials} and \textsc{Exams} where we only need to handle multiple-choice questions in \crowdaq{}'s implementation, a major challenge for \textsc{Task} is how to meet different requirements for annotation UI from different datasets in a single framework, which we discuss next.

\section{Customizable Annotation Interface}
\label{sec:interface}
It is time-consuming for non-experts on the frontend to design annotation UIs for various datasets. At present, requesters can only reuse the UIs of very similar tasks and still, they often need to make modifications with additional tests and debugging.
\crowdaq comes with a variety of built-in resources for easily creating UIs, which we will explain using an example dataset collection project centered around confirmed COVID-19 cases and deaths mentioned in news snippets.

\subsection{Concepts}
The design of \crowdaq{}'s annotation UI is built on some key concepts.
First, every \textsc{Task} is associated with \texttt{contexts}---a list of objects of any type: \texttt{text}, \texttt{html}, \texttt{image}, \texttt{audio}, or \texttt{video}. 
It will be visible to the annotators during the entire annotation process before moving to the next \textsc{Task}, so a requester can use \texttt{contexts} to show any useful information to the annotators. Below is an example of showing notes and a target news snippet (see Fig.~\ref{fig:context} in the appendix for visualization).  \crowdaq{} is integrated with online editors that can auto-complete, give error messages, and quickly preview any changes. 

\begin{lstlisting}
"contexts": [
    {
        "label": "Note",
        "type": "html",
        "html": "<p>Remember to ...</p>",
        "id": "note"
    },
    {
        "type": "text",
        "label": "The snippet was from an article published on 2020-05-20 10:30:00",
        "text": "As of Tuesday, 144 of the state's then-294 deaths involved nursing homes or longterm care facilities.",
        "id": "snippet"
    }
],
\end{lstlisting}

Second, each \textsc{Task} may have multiple \texttt{annotations}. 
Although the number of dataset formats can be arbitrary, we observe that the most basic formats fall into the following categories: multiple-choice, span selection, and free text generation.
For instance, to emulate the data collection process used for the CoNLL-2003 shared task on named entity recognition~\cite{TjongDe03}, one could use a combination of a span selection (for selecting a named entity) and a multiple-choice question (selecting whether it is a person, location, etc.); for the process used for natural language inference in SNLI~\cite{bowman2015large}, one could use an input box (for writing a hypothesis) and a multiple-choice question (for selecting whether the hypothesis entails or contradicts the premise); for reading comprehension tasks in the question-answering (QA) format, one could use an input box (for writing a question) and a multiple-choice question (for yes/no answers; \citet{CLCKCT19}), 
a span selection (for span-based answers; \citet{RZLL16}), 
or another input box (for free text answers; \citet{kovcisky2018narrativeqa}).

%
These \texttt{annotation} types are built in \crowdaq{},\footnote{For a complete list, please refer to our documentation.} which requesters can easily use to compose complex UIs.
For our example project, we would like the annotator to select a quantity from the ``snippet'' object in the \texttt{contexts}, and then tell us whether it is relevant to COVID-19 (see below for how to build it and Fig.~\ref{fig:quant ext and typing 1} in the appendix for visualization).


\begin{lstlisting}
"annotations": [
    {
        "type": "span-from-text",
        "from_context": "snippet",
        "prompt": "Select one quantity from below.",
        "id": "quantity",
    },
    {
        "type": "multiple-choice",
        "prompt": "Is this quantity related to COVID-19?",
        "options":{
            "A": "Relevant",
            "B": "Not relevant"
        }
        "id": "relevance"
    }
]
\end{lstlisting}

Third, a collection of \texttt{annotations} can form an \texttt{annotation group} and a \textsc{Task} can have multiple of them. 
For complex \textsc{Tasks}, this kind of semantic hierarchy can provide a big picture for both the requesters and annotators.
We are also able to provide very useful features for \texttt{annotation groups}.
For example, we can put the \texttt{annotations} object above into an \texttt{annotation group}, and require 1-3 responses in this group. 
Below is its syntax, and Fig.~\ref{fig:quant ext and typing 2} in the appendix shows the result.

\begin{lstlisting}
"annotation_groups": [
    {
        "annotations": [
            {"id": "quantity", ...},
            {"id": "relevance", ...}
        ],
        "id": "quantity_extraction_typing",
        "title": "COVID-19 Quantities",
        "repeated": true, "min": 1, "max": 3
    }
],
\end{lstlisting}

\subsection{Conditions}
Requesters often need to collect some \texttt{annotations} only when certain \texttt{conditions} are satisfied. For instance, only if a quantity is related to COVID-19 will we continue to ask the type of it.
These \texttt{conditions} are important because by construction, annotators will not make mistakes such as answering a question that should not be enabled at all.

As a natural choice, \crowdaq{} has implemented \texttt{conditions} that take as input values of multiple-choice \texttt{annotations}. 
The field \texttt{conditions} can be applied to any \texttt{annotation}, which will be enabled only when the \text{conditions} are satisfied.
Below we add a multiple-choice question asking for the type of a quantity {\em only if} the annotator has chosen option ``A: Relevant'' in the question whose ID is ``relevance'' (see Fig.~\ref{fig:condition example} in the appendix).

\begin{lstlisting}
"annotations": [
    {"id": "quantity", ...},
    {"id": "relevance", ...},
    {
        "id": "typing",
        "type": "multiple-choice",
        "prompt": "What type is it?",
        "options":{
            "A": "Number of Deaths",
            "B": "Number of confirmed cases",
            "C": "Number of hospitalized",
            ...
        },
        "conditions":[
            {
                "id": "relevance",
                "op": "eq",
                "value": "A"
            }
        ]
    }
],
\end{lstlisting}

\crowdaq{} actually supports any boolean logic composed by ``AND,'' ``OR,'' and ``NOT.'' Below is an example of $\neg (Q1=A \vee Q2=B)$.

\begin{lstlisting}
"conditions": [
	{
		"op": "not", "arg": {
			"op": "or", "args":[
				{"id": "Q1","op": "eq","value": "A"},
				{"id": "Q2","op": "eq","value": "B"}
			]
		}
	}
]
\end{lstlisting}

\begin{figure*}[t!]
    \centering
    \includegraphics[width=\textwidth]{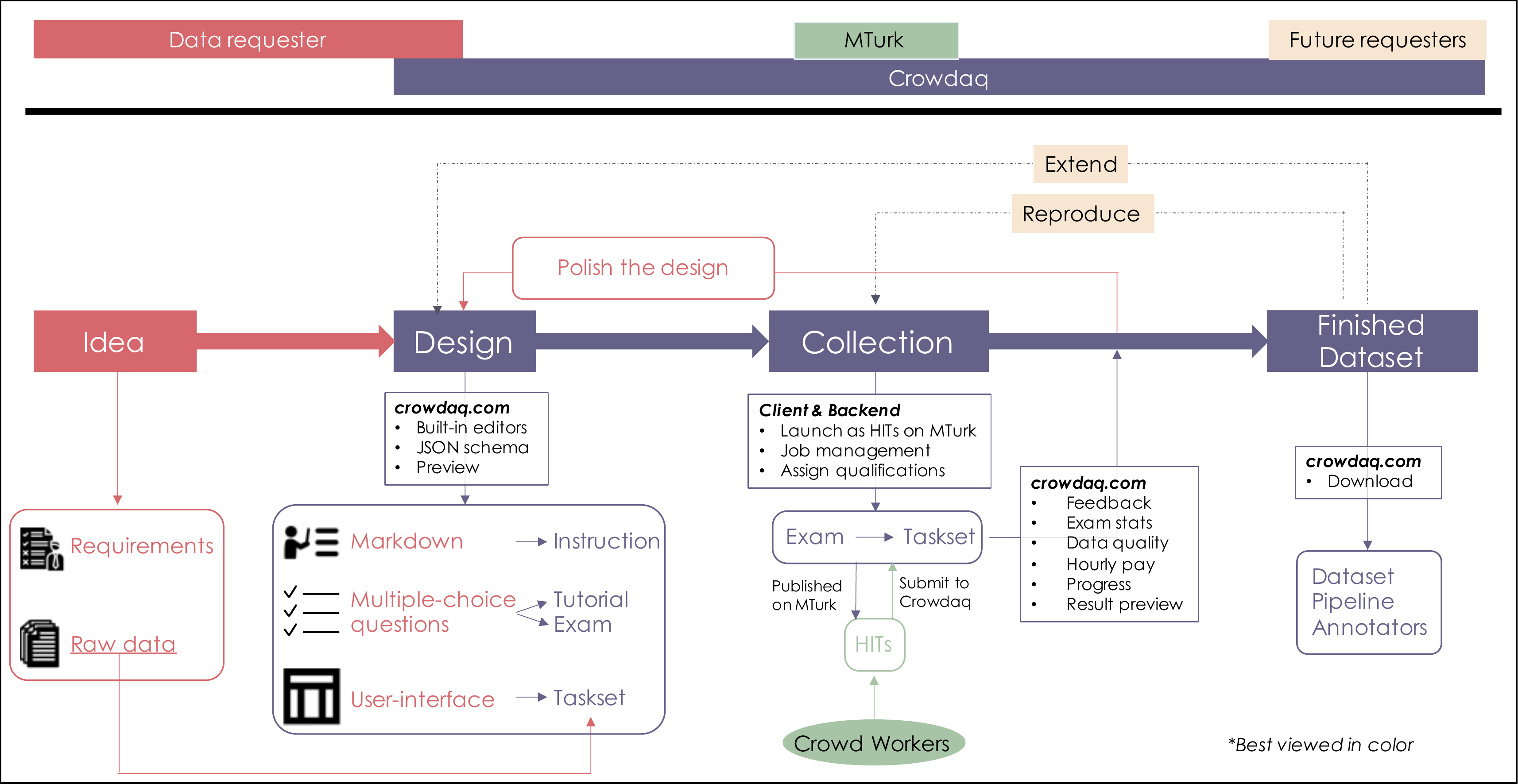}
    \caption{Data collection using \crowdaq{} and MTurk. Note that this is a general workflow and one can use only part of it, or use it to build even more advanced workflows.}
    \label{fig:usage}
\end{figure*}
\subsection{Constraints}
\label{sec:const}
An important quality control mechanism is to implement {constraints} for an annotator's work such that only if the {constraints} are satisfied will the annotator be able to submit the instance (and get paid). 
An implicit constraint in \crowdaq{} is that all \texttt{annotations} should be finished except for those explicitly specified as ``optional.'' 

For things that are repeated, \crowdaq{} allows the requester to specify 
the min/max number of repetitions. This corresponds to scenarios where, for instance, we know there is at least 1 quantity (min=1) in a news snippet or we want to have exactly two named entities selected for relation extraction (min=max=2). We have already shown usages of this when introducing \texttt{annotation group}, but the same also applies to text span selectors.

\crowdaq{} also allows requesters to specify a regular expression constraint. For instance, in our COVID-19 example, when the annotator selects a text span as a quantity, we want to make sure that the span selection does not violate some obvious rules. 
To achieve this, we define \texttt{constraints} as a list of requirements and all of them must be satisfied; if any one of them is violated, the annotator will receive an error message specified by the \texttt{description} field and also not able to submit the work. 

In addition, users can specify their own constraint functions via an API. Please refer to our documentation for more details. 
\begin{lstlisting}
"annotations":[
    {
        "id": "quantity",
        ...,
        "constraints": [
            {
                "description": "The quantity should only start with digits or letters.",
                "regex": "^[\\w\\d].*$",
                "type": "regex"
            },
            {
                "description": "The quantity should only end with digits, letters, or %.",
                "regex": "^.*[\\w\\d%]$",
                "type": "regex"
            },
            {
                "description": "The length of your selection should be within 1 and 30.",
                "regex": "^.{1,30}$",
                "type": "regex"
            }
        ]
    },
    ...
]
\end{lstlisting}







\subsection{Extensibility}
As we cannot anticipate every possible UI requirement, we have designed \crowdaq{} to be extensible. 
In addition to a suite of built-in \texttt{annotation} types, \texttt{conditions}, and \texttt{constraints}, users can write their own components and contribute to \crowdaq{} easily. All these components are separate Vue.js\footnote{\url{https://vuejs.org/}} components and one only needs to follow some input/output specifications to extend \crowdaq{}.

\section{Usage}
\label{sec:usage}

We have already deployed \crowdaq{} at \url{https://www.crowdaq.com} with load balancing, backend cluster, relational database, failure recovery, and user authentication. Data requesters can simply register and enjoy the convenience it provides. For users who need to deploy \crowdaq{}, we provide a Docker compose configuration so that they can bring up a cluster with all the features with one single command. Users will need to have their own domain name and HTTPS certificate in that case in order to use \crowdaq{} with MTurk.

Figure~\ref{fig:usage} shows how a requester collects data using \crowdaq{} and MTurk. The steps are: (1) identify the requirements of an application and find the raw data that one wants to annotate; (2) design the data collection pipeline using the built-in editors on \crowdaq{}'s website, including the Markdown \textsc{Instruction}, \textsc{Tutorial}, \textsc{Exam}, and \textsc{Interface}; (3) launch the \textsc{Exam} and \textsc{Task Set} onto MTurk and get crowd annotators to
work on them; (4) if the quality and size of the data have reached one's requirement, publish the dataset. 
We have color-coded those components in Fig.~\ref{fig:usage} to show the responsibilities of the data requester, \crowdaq{}, MTurk, and future requesters who want to reproduce or extend this dataset.
We can see that \crowdaq{} significantly reduces the effort a data requester needs to put in implementing all those features.




We have described how to write \textsc{Instructions}, \textsc{Tutorials}, and \textsc{Exams} (Sec.~\ref{sec:pipeline}) and how to design the annotation UI (Sec.~\ref{sec:interface}).
Suppose we have provided 20 \textsc{Exam} questions for the COVID-19 project. Before launching the \textsc{Exam}, we need to configure the sample size of the \textsc{Exam}, the passing score, and total number of chances (e.g., every time a participant will see a random subset of {10} questions, and to pass it, one must get a score higher than {80\%} within {3} chances). 
This can be done using the web interface of \crowdaq{} (see Fig.~\ref{fig:create exam} in the appendix). 

It is also very easy to launch the \textsc{Exam} to MTurk. \crowdaq{} comes with a client package that one can run from a local computer (Fig.~\ref{fig:launch exam} in the appendix). The backend of \crowdaq{} will do the job management, assign qualifications, and provide some handy analysis of how well participants are doing on the exam, including the score distribution of participants and analysis on each individual questions (Fig.~\ref{fig:exam stats}).

The semantic difference between \textsc{Exams} and \textsc{Task Sets} is handled by the backend of \crowdaq{}. From MTurk's perspective, \textsc{Exams} and \textsc{Task Sets} are both \textsc{ExternalQuestions}.\footnote{\textsc{ExternalQuestion} is a type of HITs on MTurk.}
Therefore, the same client package shown in Fig.~\ref{fig:launch exam} can also be used to launch a \textsc{Task Set} to MTurk. \crowdaq{}'s backend will receive the annotations submitted by crowd workers; the website will show the annotation progress and average time spent on each \textsc{Task}, and also provide quick preview of each individual annotations. If the data requester finds that the quality of annotations is not acceptable, the requester can go back and polish the design.

When data collection is finished, the requester can download the annotation pipeline and list of annotators from \crowdaq, and get information about the process such as the average time spent by workers on the task (and thus their average pay rate).  Future requesters can then use the pipeline as the starting point for their projects, if desired, e.g., using the same \textsc{Exam}, to get similarly-qualified workers on their follow-up project.

Although Fig.~\ref{fig:usage} shows a complete pipeline of using \crowdaq{} and MTurk, \crowdaq{} is implemented in such a way that data requesters have the flexibility to use only part of it. For instance, one can only use \textsc{Instruction} to host and render Markdown files, only use \textsc{Exam} to test annotators, or only use \textsc{Task Set} to quickly build annotation UIs. 
One can also create even more advanced workflows, e.g., using multiple \textsc{Exams} and filtering annotators sequentially \cite[e.g., Gated Instruction;][]{LSBLLW16}, creating a second \textsc{Task Set} to validate previous annotations, or splitting a single target dataset into multiple components, each of which has its own \textsc{Exam} and \textsc{Task Set}. 
In addition, data collection with in-house annotators can be done on \crowdaq{} directly, instead of via MTurk.
For instance, data requesters can conveniently create a contrast set \cite{GABBBCDDEGetal20} on \crowdaq{} by themselves. 

We have put more use cases into the appendix, including DROP \cite{DWDSSG19}, MATRES \cite{NingWuRo18}, TORQUE \cite{NWHPGR20}, VQA-E \cite{Li2018VQAEEE}, and two ongoing projects.




\section{Related Work}
\paragraph{Crowdsourcing Platforms} The most commonly used platform at present is MTurk, and the features \crowdaq{} provides are overall complementary to it.
\crowdaq{} provides integration with MTurk, but it also allows for in-house annotators and any platform that provides crowdsourcing service. 
Other crowdsourcing platforms, e.g., CrowdFlower/FigureEight,\footnote{\url{https://www.figure-eight.com/}} Hive,\footnote{\url{https://thehive.ai/}} and Labelbox,\footnote{\url{https://labelbox.com/}} also have automated qualification control as \crowdaq{}, but they do not separate the format of an exam from the format of a main task; therefore it is impossible to use its built-in qualification control for non-multiple-choice tasks like question-answering. In addition, \crowdaq{} provides huge flexibility in annotation UIs as compared to these platforms. Last but not least, \crowdaq{} is open-source and can be used, contributed to, extended, and deployed freely. 

\paragraph{Customizable UI} To the best of our knowledge, existing works on customizable annotation UI, e.g., MMAX2\footnote{\url{http://mmax2.net/}} \cite{MullerSt06}, PALinkA \cite{Orasan03}, and BRAT\footnote{\url{https://brat.nlplab.org/about.html}} \cite{SPTOAT12}, were mainly designed for in-house annotators on classic NLP tasks, and their adaptability and extensibility are limited. 

\paragraph{AMTI} is a command line interface 
for interacting with MTurk,\footnote{\url{https://github.com/allenai/amti}}
while \crowdaq{} is a website providing one-stop solution including instructions, qualification tests, customizable interfaces, and job management on MTurk. AMTI also addresses the reproducibility issue by allowing HIT definitions to be tracked in version control, while \crowdaq{} addresses by standardizing the workflow and automated qualification control.

\paragraph{Sprout} by \citet{BraggWe18} is a {\em meta-framework} similar to the proposed workflow. They focus on {\em teaching} crowd workers, while \crowdaq{} spends most of engineering effort to allow requesters specify the workflow {\em declaratively} without being a frontend or backend expert. 

%

\section{Conclusion}
Efficient data collection at scale is important for advancing research and building applications in NLP. 
Existing workflows typically require multiple annotators, which 
introduces overhead in building annotation UIs and training and filtering annotators.
\crowdaq{} is an open-source online platform aimed to reduce this overhead and improve reproducibility via customizable UI components, automated qualification control, and easy-to-reproduce pipelines.
The rapid modeling improvements seen in the last few years need a commensurate improvement in our data collection processes, and we believe that \crowdaq is well-situated to aid in easy, reproducible data collection research.


\bibliographystyle{acl_natbib}
\bibliography{ccg-long,cited-long,emnlp2020}

\clearpage
\newpage
\appendix
\section{Additional Use Cases}
In this appendix we describe additional use cases of \crowdaq{}. We mainly describe the task definitions and what types of annotation UIs they require. For more information, please refer to the section for examples in our documentation at \url{https://www.crowdaq.com/}.

\subsection{DROP}
DROP\footnote{\url{https://allennlp.org/drop}} \cite{DWDSSG19} is a reading comprehension dataset which focuses on performing discrete reasoning/operation over multiple spans in the context to get the final answer. The input contexts were sampled from Wikipedia with high frequency of numbers. Annotators from MTurk were asked to write 12 questions per context. The answers can be a number (free-form input), a date (free-form input) or a set of spans (from the context).

Following the original DROP dataset collection guidelines we create a similar annotation task on \crowdaq. 
We pose \textsc{Exam} questions like:
\begin{compactitem}
\item Which of the following is (not) a good question that you will write for this task?,
\end{compactitem}
where correct options are those questions that require discrete reasoning and incorrect ones are those that do not require such type of reasoning (e.g., questions that require predicate-argument structure look-ups). 

{On \crowdaq, we define an \texttt{annotation group} that require at least 12 repetitions (min=12) of the following: an input box for writing a question, a multiple-choice question for selecting the type of answer (i.e., a number that does not appear in text, a date, a set of spans from the context), and then depending on the type (fulfilled by using \texttt{conditions}), we will show an input box, a datetime collector, or a span selector, all from the built-in \texttt{annotation} types in \crowdaq{}.
}

The original DROP interface had a lot of complex constraints to ensure the quality of collected data, which can easily be implemented in \crowdaq{} using the customized constraints API described at the end of Sec.~\ref{sec:const}. 
\begin{compactitem}
    \item A constraint over a set of annotation objects (number, date or spans) to ensure that the worker has provided an answer in at least one of the annotation objects.
    \item A constraint on question annotation object to allow only how-many type of questions when the answer is a number.
    \item A task-level constraint to ensure that a worker does not repeat a previously written question within the same task.
\end{compactitem}



\subsection{MATRES}
MATRES\footnote{\url{https://github.com/qiangning/MATRES}} \cite{NingWuRo18} is a dataset for temporal relation extraction. The task is to determine the temporal order of two events, i.e., whether some event happened \emph{before}, \emph{after}, or \emph{simultaneously with} another event. MATRES took the articles and all the events annotated in the TempEval3 dataset \cite{ULADVP13}, and then used CrowdFlower to label relations between these events. Specifically, crowd workers were first asked to label the axis of each event according to the multi-axis linguistic formalism developed in \citet{NingWuRo18}, and then provide a label for every pair of events that are on the same axis.

Similar to \citet{NingWuRo18}, we split the task into two steps: axis annotation and relation annotation. The original UI for axis annotation would show a sentence with only one event highlighted (we can use an \texttt{html} \texttt{context} in \crowdaq{}), and ask for a label for the axis (we can use the multiple-choice \texttt{annotation} type). As for relation annotation, the original UI would show two events on the same axis (we can use an  \texttt{html context}) and ask for a label for the temporal ordering (multiple-choice \texttt{annotation}). Both steps are readily supported by \crowdaq{}.

 Moreover, \crowdaq{} has more advanced features to even improve the original UIs designed in \citet{NingWuRo18}. For instance, when showing two events on the same axis, we can ask if the annotator agrees with the same-axis claim, and only if the annotator agrees with it, we ask for a relation between them. This will reduce axis errors propagated from the axis annotation step.

\subsection{TORQUE}
TORQUE \cite{NWHPGR20} is a reading comprehension dataset composed of questions specifically on temporal relations. For instance, given a passage, ``{\em Rescuers searching for a woman trapped in a landslide said they had found a body},'' typical questions in TORQUE are: 
\begin{compactitem}
\item What happened before a woman was trapped?
\item what happened while a woman was trapped?
\end{compactitem}

Annotators of TORQUE are required to identify events from a passage, ask questions that query temporal relations, and answer those questions correctly.
The original instruction and qualification exam for TORQUE are publicly available at \url{https://qatmr-qualification.github.io/}. Since its qualification exam is already in the format of multiple-choice questions, we can easily transfer it to \crowdaq{}.

The original UI for the main annotation process of TORQUE is also available at \url{https://qatmr.github.io/}. On each passage, it has the following steps (corresponding components of \crowdaq{} in parentheses): (1) label all the events in the passage (span selector); (2) answer 3 pre-defined, warm-up questions (span selector); (3) write a new question that queries temporal relations (input box); (4) answer the question (span selector) using those events labeled in Step 1 (customized \texttt{constraints}); (5) repeat Step 3 \& 4 for at least 12 times (min=12 in the \texttt{annotation group}). \crowdaq{} supports all these features.

\subsection{VQA-E}

VQA-E\footnote{\url{https://github.com/liqing-ustc/VQA-E}} \cite[Visual Question Answering with Explanation;][]{Li2018VQAEEE} is a dataset for training and evaluating models that generate explanations that justify answers to questions about given images. Besides constructing such a dataset, \crowdaq{} can also be used to evaluate the plausibility of an explanation (i.e., whether a generated explanation supports the answer in the context of the image), and its visual fidelity (i.e., whether the explanation is grammatical, but mentions content unrelated to the image---anything that is not directly visible and is unlikely to be present in the scene in the image). 

We use \crowdaq{}'s \texttt{contexts} with the type \texttt{image} to display a given image, the \texttt{html} context to display a question and answer, and \crowdaq{}'s \texttt{annotation} of the type \textit{multiple-choice} to ask the following prompts: 
\begin{compactitem}
\item Does the explanation support the answer?
\item Is the explanation grammatically correct?,
\end{compactitem}
We then use \crowdaq{}'s \texttt{condition} to add another multiple-choice question: 
\begin{compactitem}
\item Does the explanation mention a person, object, location, action that is unrelated to the image?,
\end{compactitem}
that is shown only if the annotator has judged the explanation to be grammatical in the previous step. Conditioning helps us differentiate between ungrammaticality and visual ``infidelity.'' Finally, as an alternative way of measuring fidelity, we use the annotation with the type \textit{multi-label} to display the following prompt:
\begin{compactitem}
\item Select nouns that are unrelated to the image, 
\end{compactitem}
where nouns are extracted from the explanation, and the difference between \textit{multiple-choice} and \textit{multi-label} is that the latter allows for more than one options to be selected. However, in the \textsc{Exam}, we teach annotators to select such nouns with the multiple-choice questions: 
\begin{compactitem}
\item Is the noun $\langle$\textit{insert\textunderscore noun}$\rangle$ related to the image?
\end{compactitem}
Because judging explanation plausibility and fidelity is difficult and subjective, \crowdaq{}'s \textsc{Tutorial} and \textsc{Exam} are of a great value. 

\subsection{Answering Information-Seeking Questions about NLP Papers}
This is an ongoing dataset creation project aiming to collect a question answering dataset about NLP papers, where the questions written by real readers of NLP papers with domain expertise who have read only the titles and the abstracts of research papers and want to obtain information from the full text of the paper. In this case study, we focused on the more challenging task of obtaining answers, assuming that the questions are available. We used \crowdaq{} to design a \textsc{Tutorial} for instructing workers, and an \textsc{Exam} for qualifying them.

The task involves two steps: The first is identifying \textit{evidence} for the question, which can be a passage of text, a figure, or a table in the paper that is sufficient to answer the question. The second step is providing an \textit{answer}, which can be text that is either selected from the paper or written out free form, or a boolean (Yes/No). Some questions may be identified as being unanswerable, and do not require evidence and answers. We used the \textsc{Tutorial} and \textsc{Exam} features in \crowdaq{} to teach the workers and evaluate them on the following aspects of the task:
\begin{compactitem}
    \item \textbf{Identifying sufficient evidence} Quite often papers have several passages that provide information relevant to the question being asked, but they do not always provide all the information needed to answer them. We identified such passages that are relevant to real questions, and made \textsc{Tutorial} and \textsc{Exam} questions of the form, ``Is this evidence sufficient to answer the question?''
    \item \textbf{Preference of text over figures or tables} The task requires selecting figures or tables in NLP papers as evidence only if sufficient information is not provided by the text in the paper. To teach the workers this aspect of the task, we made multiple-choice questions showing a figure or a table from a paper, and some text referring it, and asked the workers questions of the form, ``Given this chunk of text, and this figure from the same paper, what would be good evidence for the question? (A) Just the figure (B) Just the text (C) Both (D) Neither''.
    \item \textbf{Answer type} Since the task has multiple answer types, including extractive (span selection) and abstractive (free form) answers, it is important to teach the workers when to choose each type. We made multiple-choice questions with examples showing potential span selections, comparing them with potential free-form answers, asking the workers to choose the correct option for each case.
\end{compactitem}

\subsection{Acceptability Judgement Tasks}
This is another one of the ongoing projects on \crowdaq{}.
\textit{Acceptability judgements} are a common tool in linguistics for evaluating whether a given piece of text is grammatical or semantically meaningful~\cite{chomsky2002syntactic,t2016empirical}.
Popular approaches for performing acceptability judgements include having annotators to make a Boolean evaluation of whether or not the text is acceptable, as well as forcing annotators to pick from a pair of sentences which is the most acceptable.
Both of these approaches can be easily formulated as a sequence of multiple choice questions.
Although multiple choice surveys are simple to design and deploy on many crowdsourcing platforms, \crowdaq{} users have found some features particularly useful.

First, the ability to easily design \textsc{Exams} to qualify users.
Due to their simplicity, multiple-choice questions are easily gamed by crowd workers using bots or by answering questions randomly.
In a pilot study, one requester on \crowdaq{} found that over 66\% of participants in their acceptability judgement survey were bad actors if they directly launch the task on MTurk.
By setting a performance threshold on the \textsc{Exam} these actors were automatically disqualified from participating in the \textsc{Task} without the user needing to setup and manage custom MTurk Qualifications. 
Although commercial crowd sourcing platforms provide similar qualification control, they have paywalls, are not open-sourced, and are not as flexible as \crowdaq{} (for instance, one can use only the qualification feature of \crowdaq{}, while commercial platforms require using and paying the entire pipeline).

\begin{figure}[!h]
    \centering
    \includegraphics[width=.8\linewidth]{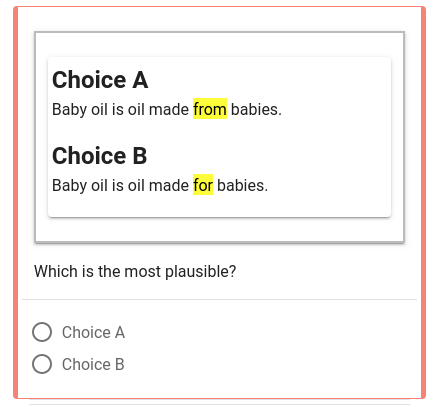}
    \caption{Illustration of an acceptability judgement task deployed on \crowdaq{}. Because \texttt{contexts} can contain \texttt{html}, this \crowdaq{} user was easily able to highlight relevant spans of text for crowd workers using \texttt{<mark>} tags. }
    \label{fig:acceptability}
\end{figure}

Second, the flexibility \crowdaq{} allows when specifying contexts.
In one case, a user received multiple requests from crowd workers to visualize which tokens differed between pieces of texts in order to increase the speed at which they were able to annotate longer pieces of text.
Since contexts allow the insertion of arbitrary HTML, this user was able to easily accommodate this request by inserting \texttt{<mark>} tags around the relevant tokens.
An illustration of one of their questions is provided in Fig.~\ref{fig:acceptability}.

\section{Screenshots}
\label{sec:appendix}
In this appendix we show screenshots corresponding to the JSON specifications described in Sec.~\ref{sec:pipeline} and Sec.~\ref{sec:interface}.
We also provide an overview of all these figures in Table \ref{tab:screenshot_summary}.

\begin{table*}[t]
\begin{tabular}{lp{14cm}}
\toprule
\textbf{Reference} & \textbf{Description}\\
\midrule
Figure \ref{fig:usage} & An illustration of data collection using \crowdaq{} and MTurk \qn{(this figure is in the main text)} \\
Figure \ref{fig:acceptability} & An illustration of an acceptability judgement task deployed on \crowdaq{} \\
Figure \ref{fig:tutorial} & A specification and visualization of \textsc{Tutorials} \\
Figure \ref{fig:exam} & A specification and visualization of \textsc{Exams} \\
Figure \ref{fig:context} & A visualization of how \crowdaq{} renders the \texttt{context} specified in Section \ref{sec:interface} \\
Figure \ref{fig:quant ext and typing 1} & An illustration of selecting a span from a text and answering a question about it on \crowdaq{} \\
Figure \ref{fig:quant ext and typing 2} & An example of repeated annotations on \crowdaq{} \\
Figure \ref{fig:condition example} & An example of how \crowdaq's \texttt{condition} works\\
Figure \ref{fig:constraint example} & An illustration of a violated \texttt{constraint} on \crowdaq{} \\
Figure \ref{fig:create exam} & A visualization of how to create an \textsc{Exam} on \crowdaq{} \\
Figure \ref{fig:launch exam} & An visualization of how to launch an exam to MTurk in the client package that comes with \crowdaq{} \\
Figure \ref{fig:exam stats} & Feature visualizations: (a) Distribution of participants' scores. (b) Individual scores of each participant. (c) Participants' performance on each question with quick preview of individual questions.\\
Figure \ref{fig:main task stats} & Feature visualizations: (a) Average time workers spend on the task. (b) Progress monitoring. (c) Quick preview of individual annotations.\\
\bottomrule
\end{tabular}
\caption{An overview of figures.}
\label{tab:screenshot_summary}
\end{table*}

\begin{figure*}[ht!]
    \centering
    \includegraphics[width=\textwidth]{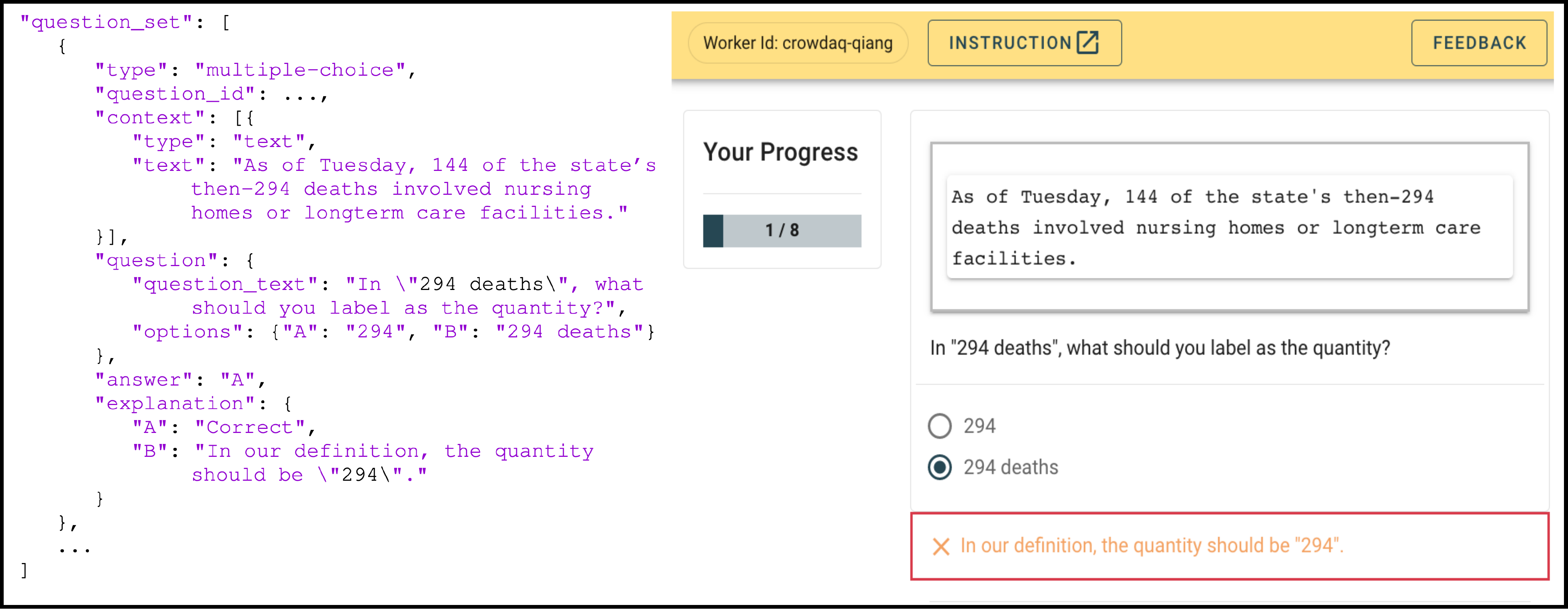}
    \caption{Specification and visualization of \textsc{Tutorials}. In this particular \textsc{Tutorial}, there are eight questions and the example participant has only made choice on one of them. Please see \url{https://beta.crowdaq.com/w/tutorial/qiang/CrowdAQ-demo} for this interface.}
    \label{fig:tutorial}
\end{figure*}

\begin{figure*}[ht!]
    \centering
    \includegraphics[width=\textwidth]{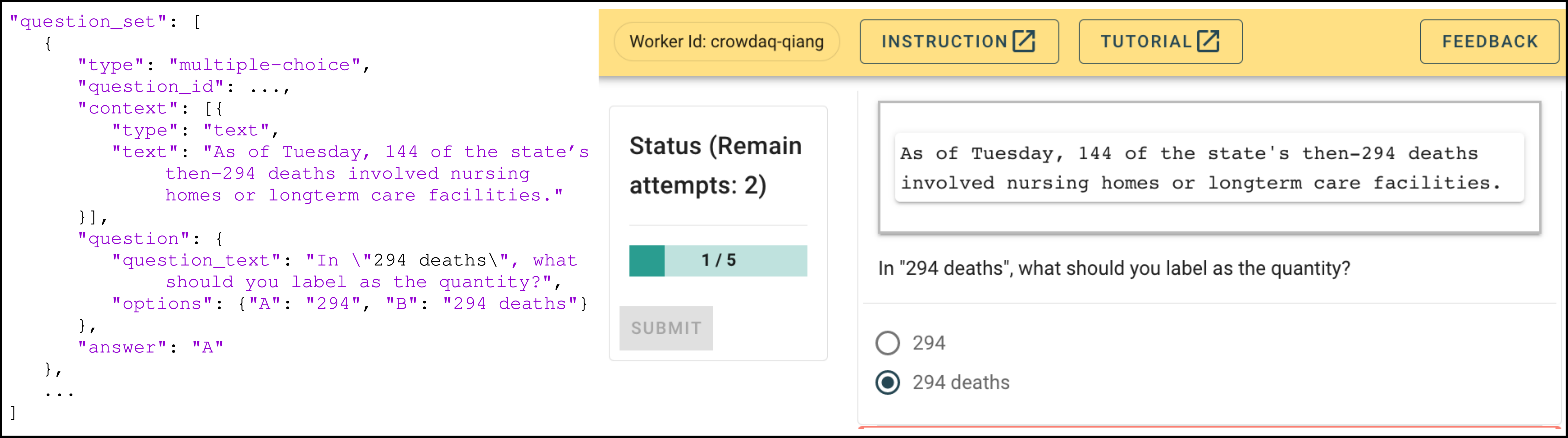}
    \caption{Specification and visualization of \textsc{Exams}. In this particular \textsc{Exam}, the requester has specified that every time a participant will see five questions randomly sampled from the pool, and every participant only has two opportunities to pass it. Please see \url{https://beta.crowdaq.com/w/exam/qiang/CrowdAQ-demo} for this interface.}
    \label{fig:exam}
\end{figure*}

\begin{figure*}[ht!]
    \centering
    \includegraphics[width=\textwidth]{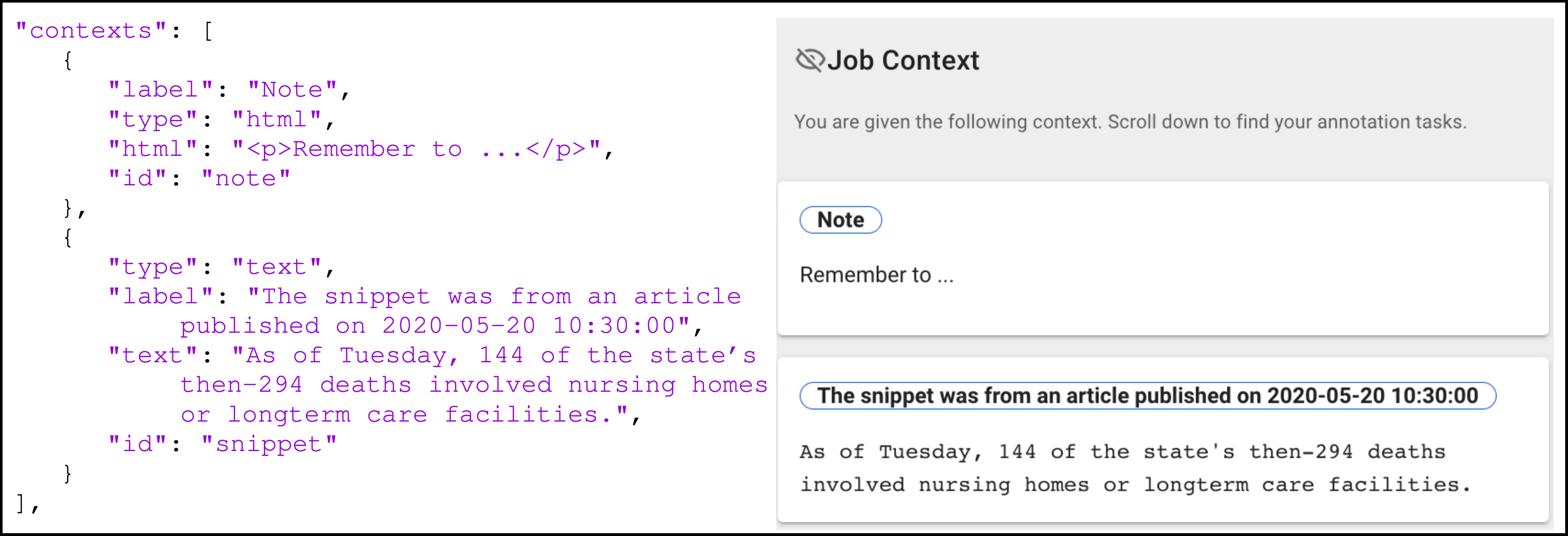}
    \caption{How \crowdaq{} renders the \texttt{context} specification in Sec.~\ref{sec:interface}. A major difference from \textsc{Tutorials} is that the participant will not see the answers. Please see \url{https://beta.crowdaq.com/w/task/qiang/CrowdAQ-demo/quantity_extraction_typing} for this interface.}
    \label{fig:context}
\end{figure*}

\begin{figure*}[ht!]
    \centering
    \includegraphics[width=\textwidth]{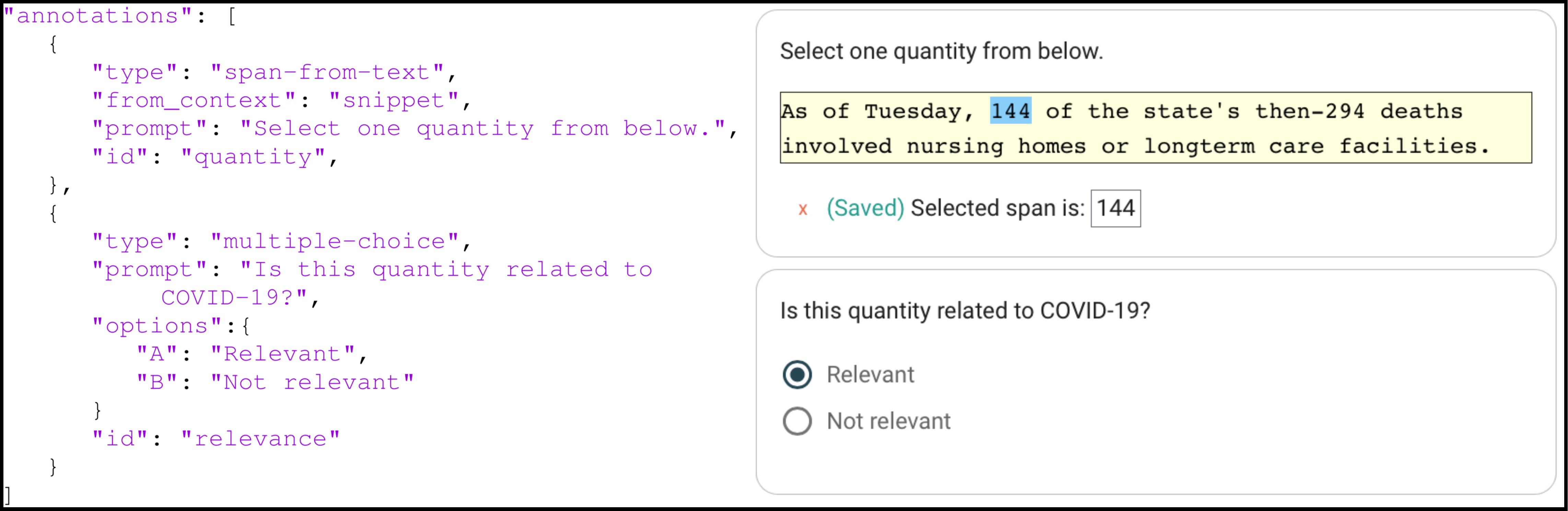}
    \caption{In this UI the annotator is asked to select a valid quantity and then choose whether it is relevant to COVID-19.}
    \label{fig:quant ext and typing 1}
\end{figure*}

\begin{figure*}[ht!]
    \centering
    \includegraphics[width=\textwidth]{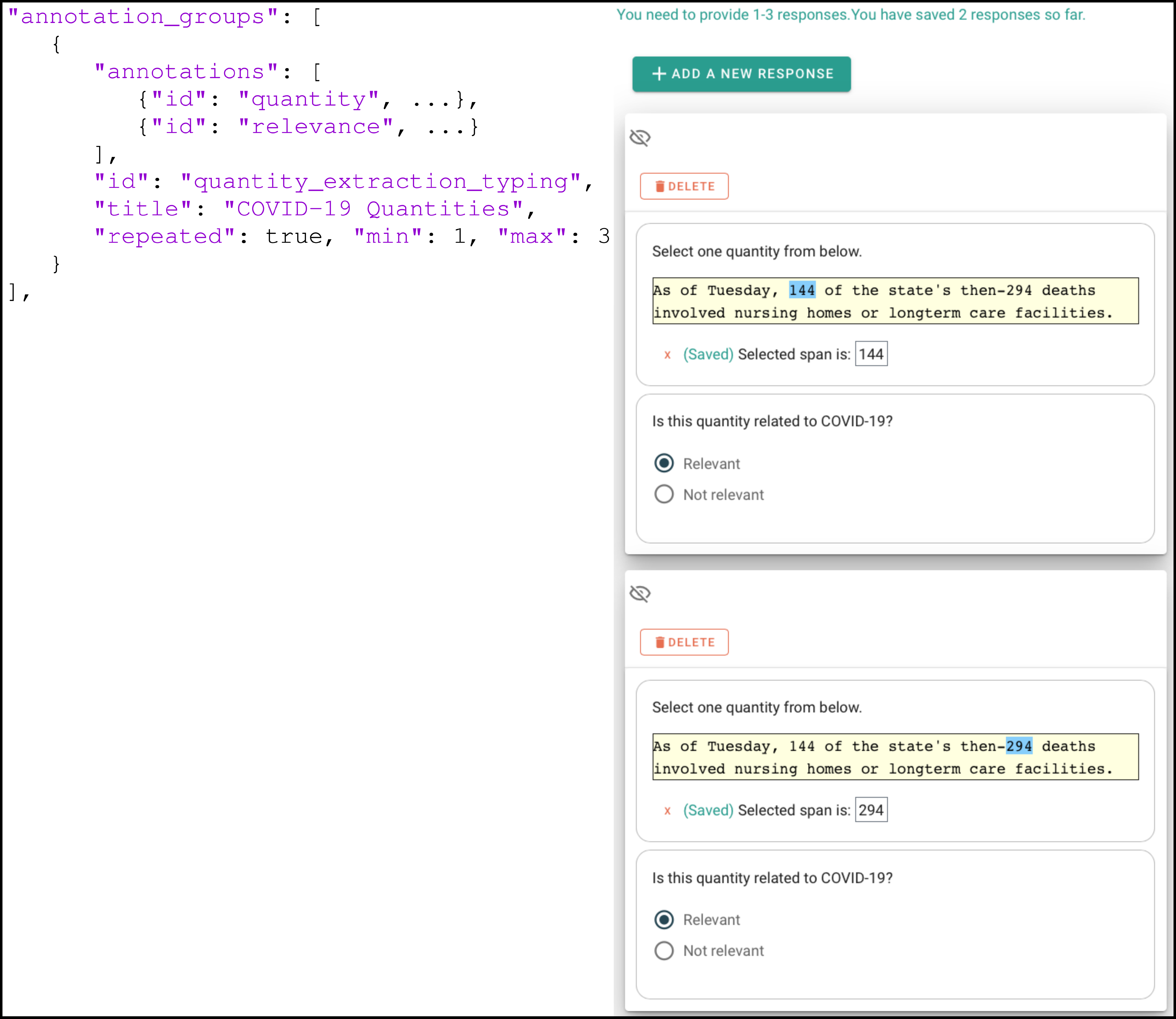}
    \caption{How to collect a group of annotations repeatedly. Note the annotation is valid only when one provides 1-3 responses because the requester specifies so. In the above we have added 2 responses.}
    \label{fig:quant ext and typing 2}
\end{figure*}

\begin{figure*}[ht!]
    \centering
    \includegraphics[width=\textwidth]{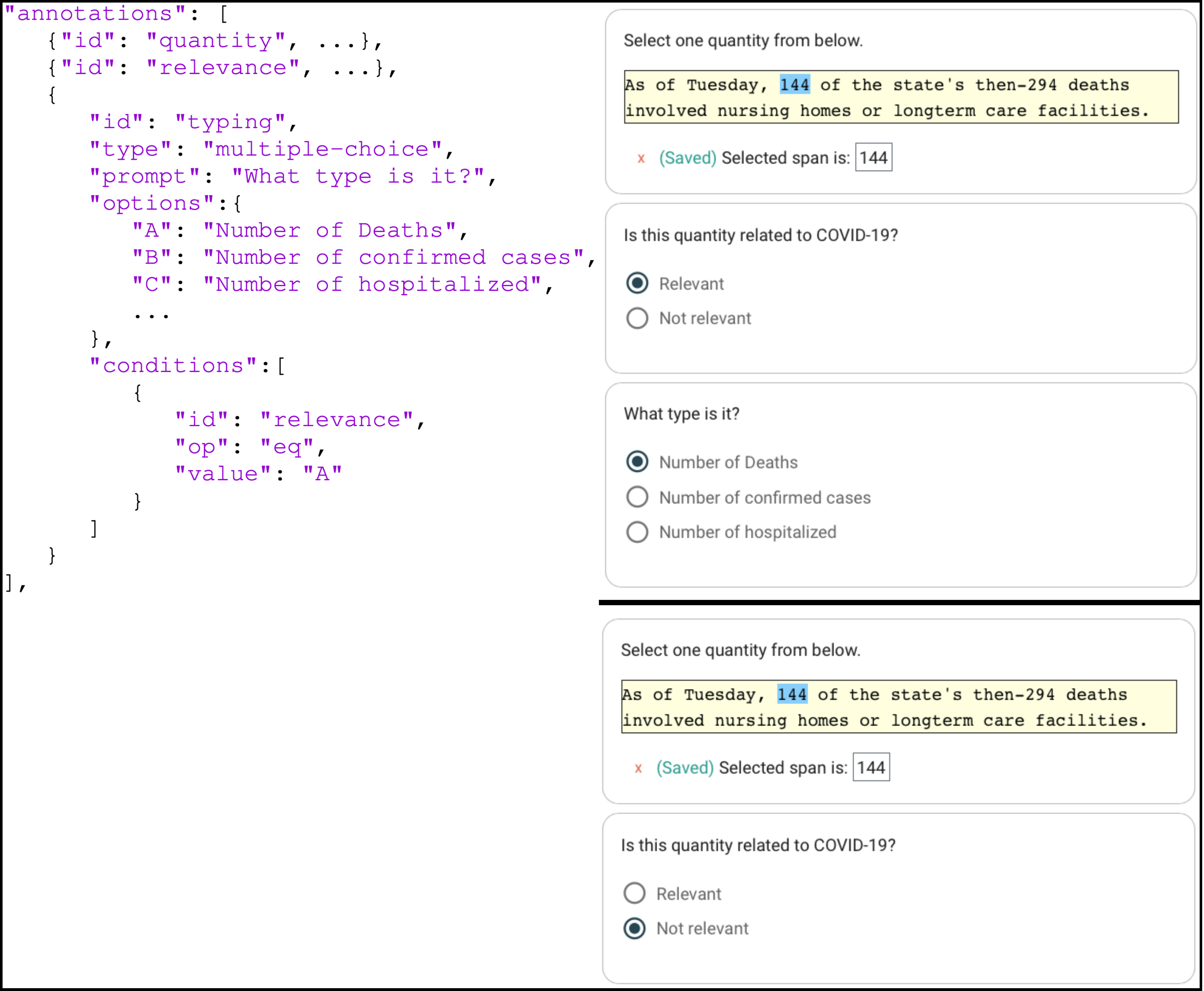}
    \caption{A new question about the type of the quantity is enabled only if we select ``Relevant'' in the previous one. Above: Enabled. Below: Disabled.}
    \label{fig:condition example}
\end{figure*}

\begin{figure*}[ht!]
    \centering
    \includegraphics[width=.6\textwidth]{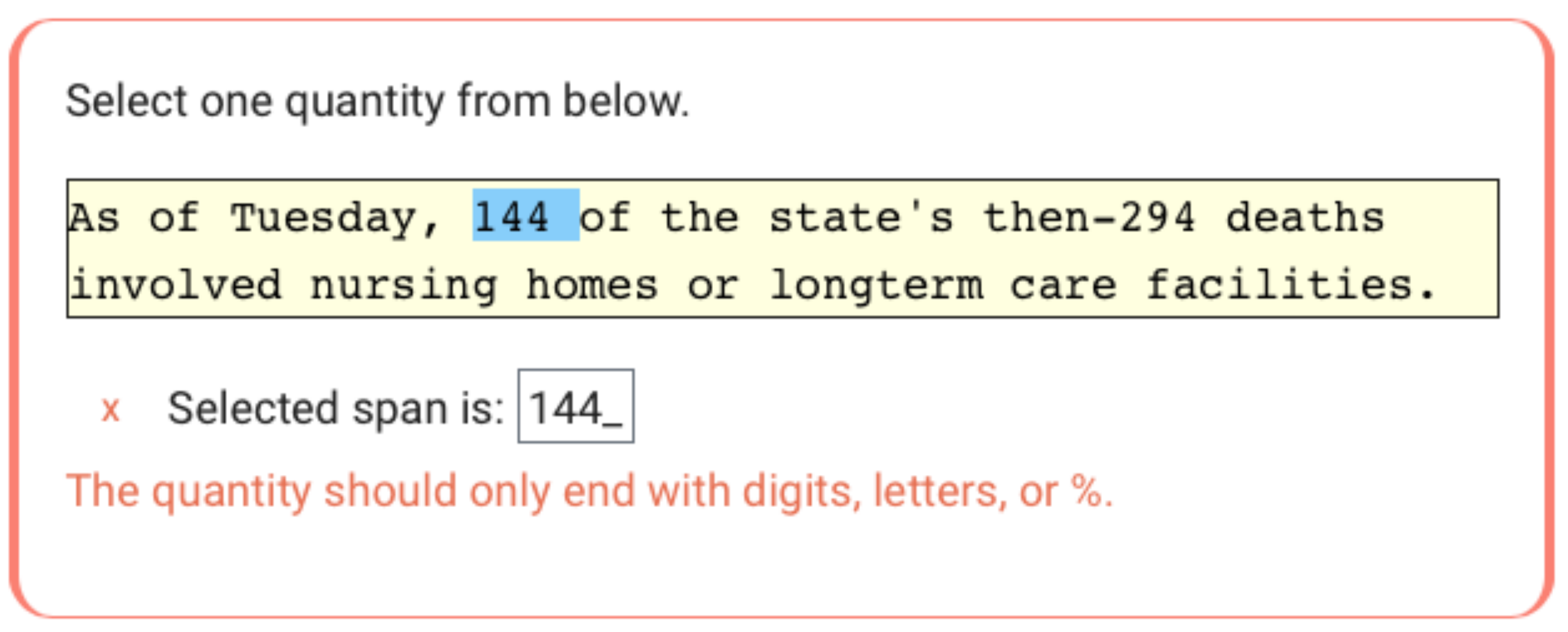}
    \caption{When any constraint is violated, the annotator will receive an error message (the text in orange) and also prohibited from proceeding (not shown here). In this screenshot, the annotator wrongly included an extra space.}
    \label{fig:constraint example}
\end{figure*}

\begin{figure*}[ht!]
    \centering
    \includegraphics[width=0.5\textwidth]{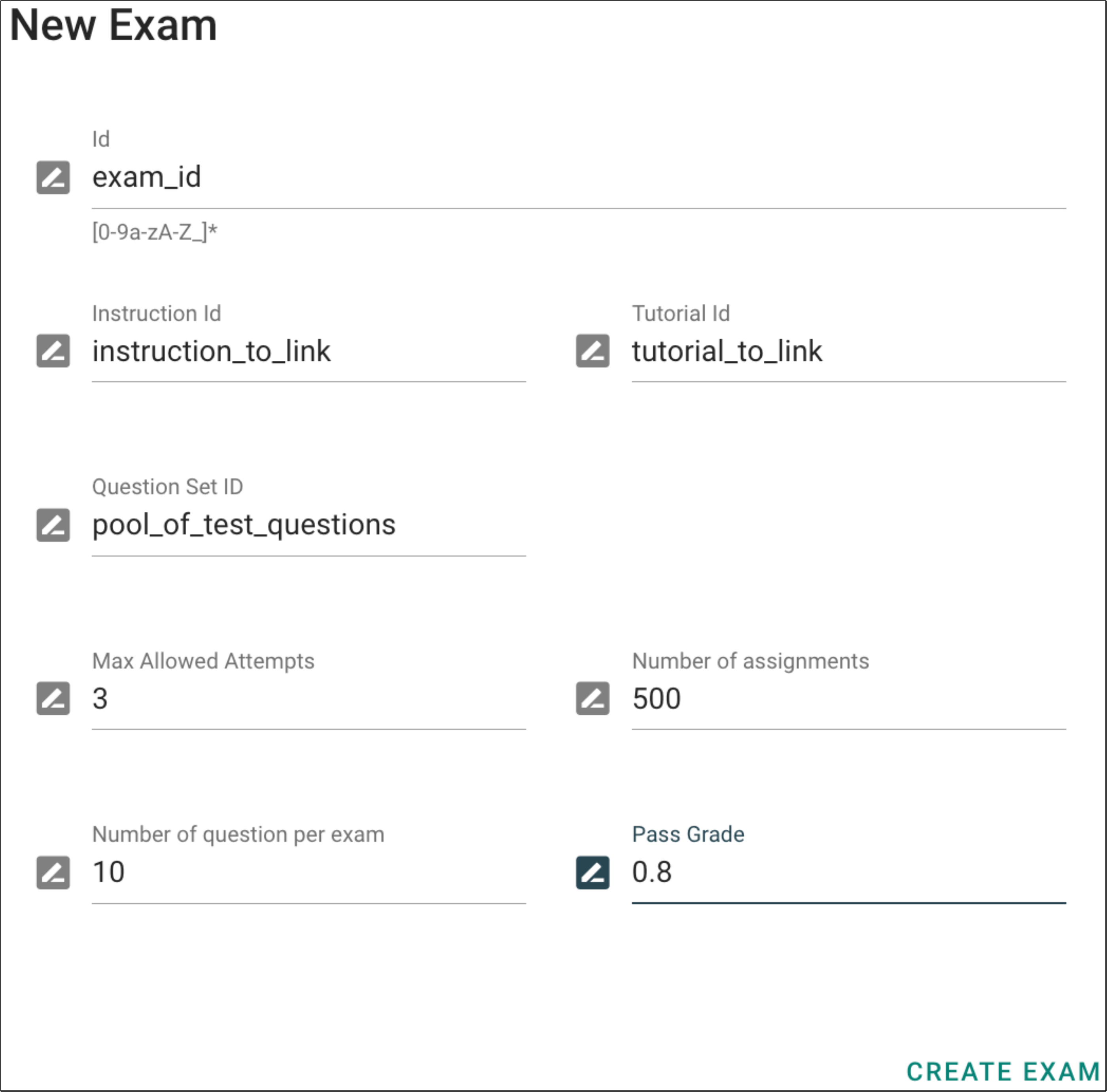}
    \caption{How to create an exam on \crowdaq{}.}
    \label{fig:create exam}
\end{figure*}

\begin{figure*}[ht!]
    \centering
    \includegraphics[width=\textwidth]{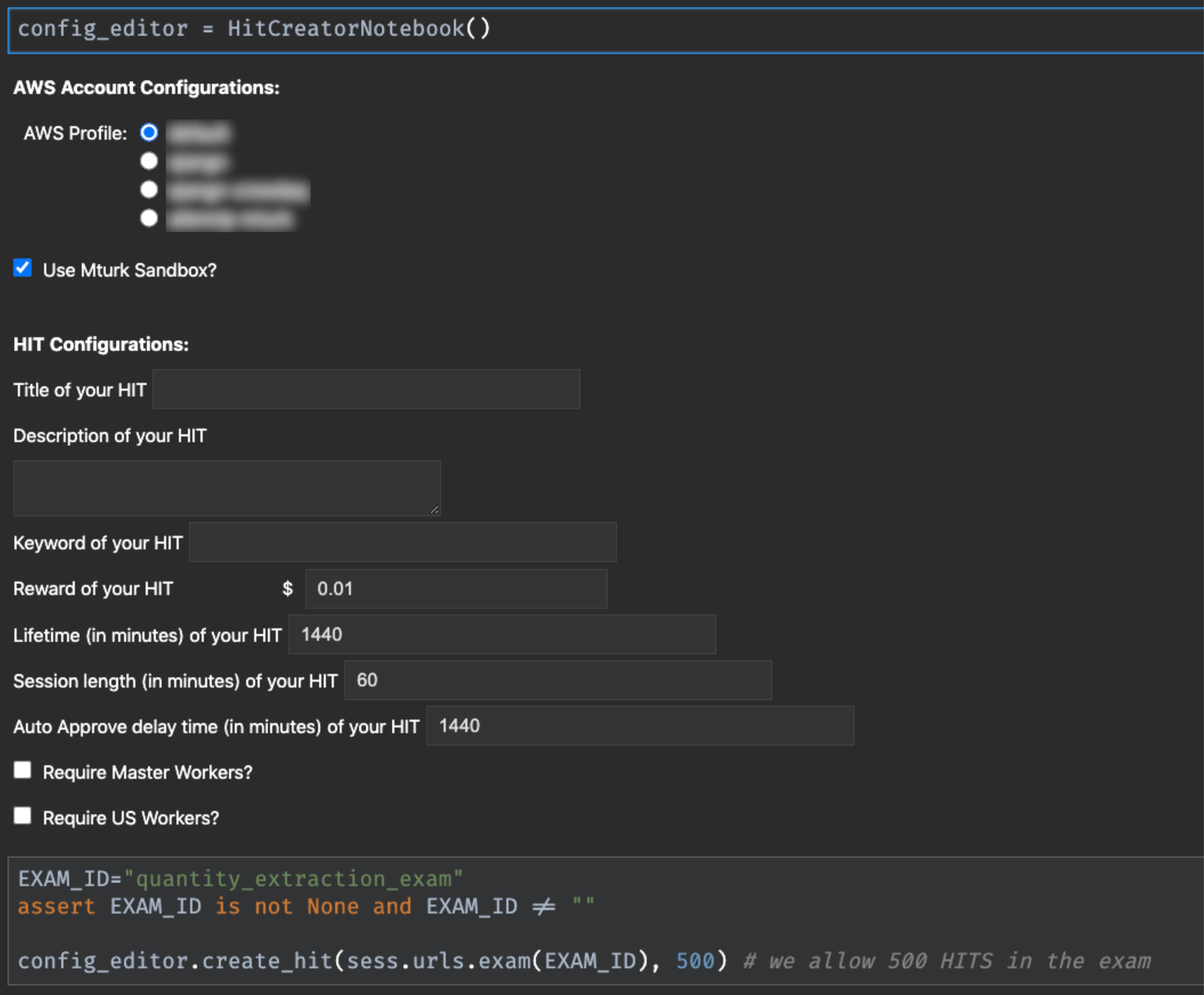}
    \caption{Launch an exam to MTurk in the client package that comes with \crowdaq{}.}
    \label{fig:launch exam}
\end{figure*}

\begin{figure*}[ht!]
    \centering
    \includegraphics[width=\textwidth]{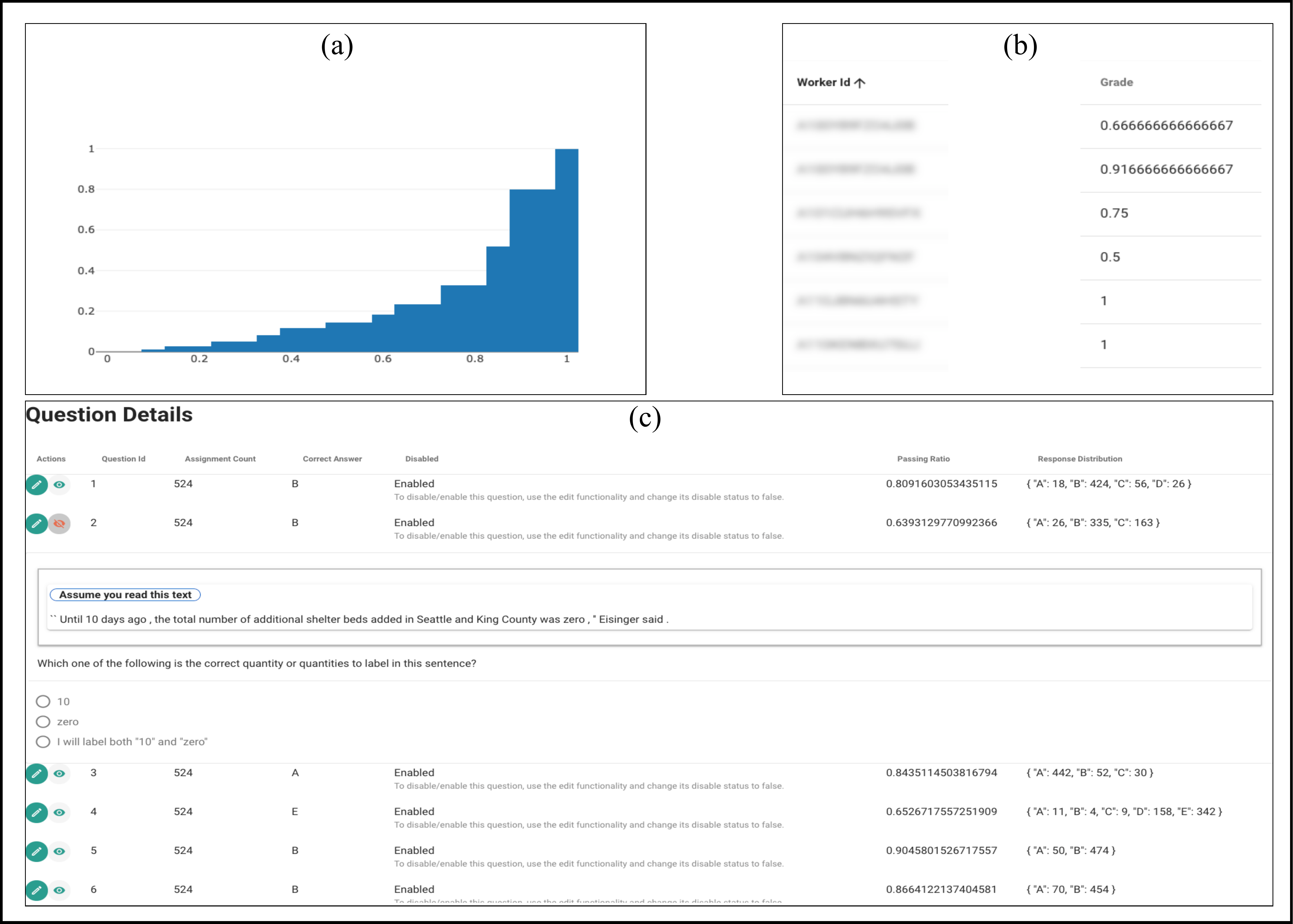}
    \caption{Some handy features that \crowdaq{} provides on \textsc{Exams}. (a) Distribution of participants' scores. (b) Individual scores of each participant. (c) Participants' performance on each question with quick preview of individual questions.}
    \label{fig:exam stats}
\end{figure*}

\begin{figure*}[ht!]
    \centering
    \includegraphics[width=\textwidth]{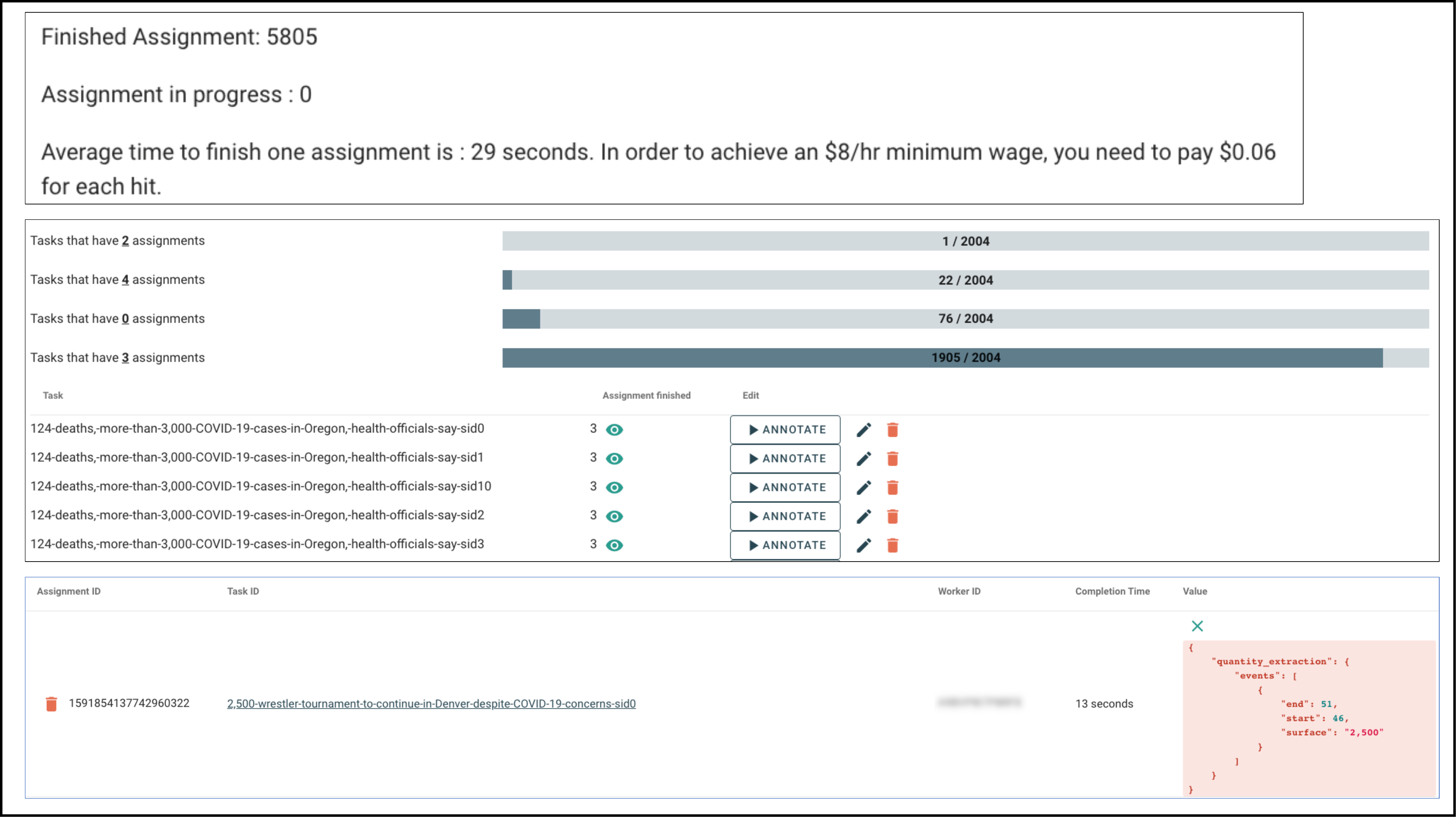}
    \caption{Some handy features that \crowdaq{} provides on \textsc{Task Sets}. Above: average time people spend on the task. Middle: progress monitoring. Below: quick preview of individual annotations.}
    \label{fig:main task stats}
\end{figure*}

\end{document}